\documentclass[twocolumn,english,prl,aps,amssymb,superscriptaddress,showpacs]{revtex4-1}
\usepackage[T1]{fontenc}
\usepackage{amsmath}
\usepackage{amssymb}
\usepackage{graphicx}

\makeatletter
\@ifundefined{textcolor}{}
{%
 \definecolor{BLACK}{gray}{0}
 \definecolor{WHITE}{gray}{1}
 \definecolor{RED}{rgb}{1,0,0}
 \definecolor{GREEN}{rgb}{0,1,0}
 \definecolor{BLUE}{rgb}{0,0,1}
 \definecolor{CYAN}{cmyk}{1,0,0,0}
 \definecolor{MAGENTA}{cmyk}{0,1,0,0}
 \definecolor{YELLOW}{cmyk}{0,0,1,0}
}

\usepackage{epsfig}
\usepackage{dcolumn}
\usepackage{bm}

\makeatother

\usepackage{babel}
\begin{document}

\title{ Noncollinear Magnetic Order Stabilized by Entangled Spin-Orbital Fluctuations 
}

\author{Wojciech Brzezicki}
\author{Jacek Dziarmaga}
\affiliation{Marian Smoluchowski Institute of Physics, Jagellonian University,
             Reymonta 4, PL-30059 Krak\'ow, Poland}

\author{Andrzej M. Ole\'{s} }
\affiliation{Marian Smoluchowski Institute of Physics, Jagellonian University,
             Reymonta 4, PL-30059 Krak\'ow, Poland}
\affiliation{Max-Planck-Institut f\"ur Festk\"orperforschung, 
             Heisenbergstrasse 1, D-70569 Stuttgart, Germany}

\date{August 15, 2012}

\begin{abstract}
Quantum phase transitions in the two-dimensional Kugel-Khomski model on 
a square lattice are studied using the plaquette mean field theory and 
the entanglement renormalization ansatz. When $3z^2-r^2$ orbitals are 
favored by the crystal field and Hund's exchange is finite, both methods 
give a noncollinear exotic magnetic order which consists of four 
sublattices with mutually orthogonal nearest neighbor and 
antiferromagnetic second neighbor spins. We derive effective frustrated 
spin model with second and third neighbor spin interactions 
which stabilize this phase and follow from spin-orbital quantum 
fluctuations involving spin singlets entangled with orbital excitations.
\end{abstract}


\maketitle

{\it Introduction.---}
Almost 40 years ago Kugel and Khomskii realized that spins and orbitals
should be treated on equal footing in Mott insulators with active 
orbital degrees of freedom \cite{Kug73}. Their model explains 
qualitatively the magnetic and orbital order in KCuF$_3$ which is
a well known example for spinon excitations in a one-dimensional (1D)
Heisenberg antiferromagnet \cite{Bella}. This archetypal compound is
usually given as an example of the spin-orbital physics \cite{Tok00},
which covers a broad class of transition metal compounds, including 
perovskite manganites \cite{Fei99}, titanates \cite{Kha00}, vanadates 
\cite{Kha01}, ruthenates \cite{Fan05}, 1D cuprates \cite{Sch12}, 
layered ruthenates \cite{Mario}, and pnictide superconductors \cite{Jan09}. 
In all these compounds strong intraorbital Coulomb repulsion $U$ 
dominates over electron hopping $t$ ($t\ll U$) and charge fluctuations 
are suppressed. On the one hand, spin degrees of freedom may separate 
from the orbitals when the coupling to the lattice is strong, as in 
LaMnO$_3$ \cite{Fei99} and recently shown to happen also in KCuF$_3$ 
\cite{Lee12}. On the other hand, the spin-orbital quantum fluctuations 
are strongly enhanced for low $S=\frac12$ spins, as in the 
three-dimensional (3D) Kugel-Khomskii (KK) model \cite{Fei97,Kha97}, 
and lead to a spin-orbital liquid phase in LaTiO$_3$ \cite{Kha00}. 
Geometrical frustration \cite{Bal10} was also suggested as a 
stabilizing mechanism for a spin-orbital liquid phase \cite{Kri05},
with examples on a triangular lattice in $e_g$ (LiNiO$_2$ \cite{Ver04}) 
and $t_{2g}$ (LiNiO$_2$ \cite{Nor08}) orbital systems. 
Frustrated spin-orbital interactions to further neighbors may also 
destabilize long-range magnetic order \cite{Nak12}.
An opposite case when {\it orbital excitations\/} determine the spin 
order was not reported until now.

The phase diagram of the 3D KK model remains controversial in the 
regime of strongly frustrated interactions --- it has been suggested 
that either spin-orbital fluctuations destabilize long-range spin 
order \cite{Fei97}, or an orbital gap opens and stabilizes spin order
\cite{Kha97}. This difficulty is typical for systems with spin-orbital 
entanglement \cite{Ole12} which may occur both in the ground state 
\cite{Ole06} and in excited states \cite{You12}. The best known 
examples are the 1D \cite{Li98} or two-dimensional (2D) \cite{Wan09} 
SU(4) models, where spin and orbital operators appear in a symmetric 
way. Instead, the symmetry in the orbital sector is much 
lower and orbital excitations measured in KCuF$_3$ \cite{Ish11} are 
expected to be inherently coupled to spin fluctuations \cite{Woh11}. 

In this Letter we present a surprising {\it noncollinear spin order\/} 
in the 2D KK model which goes beyond mean field (MF) studies 
\cite{Cha08}, and explain its origin.
So far, noncollinear spin order was obtained for frustrated exchange in 
Kondo-lattice models on square lattices, without \cite{Nor01} and with 
\cite{Lor08} orbital degeneracy, or at finite spin-orbit coupling
\cite{Fis11}. In MnV$_2$O$_4$ spinel it is
accompanied by a structural distortion and the orbital order 
\cite{Gar08}. Here we find yet a different situation --- 
when frustrated nearest neighbor (NN) exchange terms almost
compensate each other and orbitals are in ferro-orbital (FO)
state, the spin order follows from further neighbor spin interactions 
triggered by entangled spin-orbital excitations. 

%%%%%%%%%%%%%%%%%%%%%%%%%%%%%%%%%%%%%%%%%%%%%%%%%%%%%%%%%%%%%%%%%%%%%%%%%
%                            figure 1
%%%%%%%%%%%%%%%%%%%%%%%%%%%%%%%%%%%%%%%%%%%%%%%%%%%%%%%%%%%%%%%%%%%%%%%%%
\begin{figure}[t!]
\begin{center}\includegraphics[width=8.4cm]{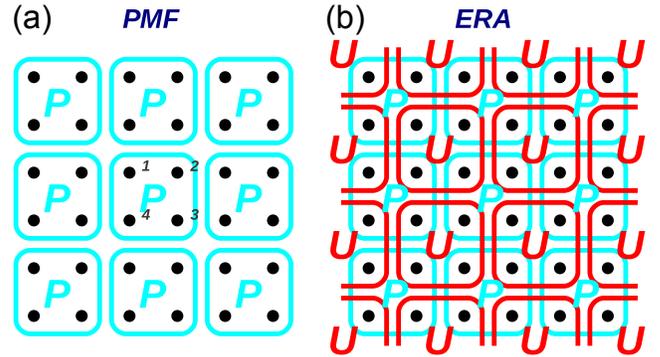}\end{center} 
\caption{(color online).
Two variational ans\"atze used in the present paper: (a) PMF, and 
(b) ERA. Black dots are lattice sites, ${\cal P}$'s are variational wave 
functions on $2\times2$ plaquettes, and ${\cal U}$'s are variational 
$2\times2$ unitary disentanglers. 
}
\label{fig:ERA} 
\end{figure}

{\it Variational approach.---}
We begin with presenting two general variational methods for 
spin-orbital systems:
(i)~the plaquette MF (PMF) ansatz, see Fig. \ref{fig:ERA}(a), and
(ii)~the entanglement renormalization ansatz (ERA) \cite{ERA}, 
see Fig. \ref{fig:ERA}(b).
In the PMF, adapted here from a similar method for the bilayer KK model 
\cite{Brz11}, one employs a variational ansatz in a form of a product of 
plaquette $2\times2$ wave functions ${\cal P}$'s \cite{plaq}. Energy is 
minimized with respect to ${\cal P}$'s to obtain the best approximation 
to the ground state. ERA is a refined version of the PMF, where the 
product of ${\cal P}$'s is subject to an additional unitary transformation, 
being a product of $2\times2$ ``disentanglers'' ${\cal U}$. They introduce 
entanglement between different plaquettes and make ERA more accurate. 

{\it Kugel-Khomskii model and methods.---} 
The perturbation theory for a Mott insulator with active $e_g$ orbitals 
in the regime of $t\ll U$ leads to the spin-orbital model \cite{Ole00}, 
with the Heisenberg SU(2) spin interactions coupled to the orbital 
operators for the holes in the $d^9$ ionic states, 
\begin{eqnarray}
\label{eq:hamik}
{\cal H} & = & -\frac{1}{2}J\sum_{\langle ij\rangle||\gamma}\left\{ 
\left(r_{1}\,\Pi_{t}^{(ij)}+r_{2}\,\Pi_{s}^{(ij)}\right)
\left(\frac{1}{4}-\tau_{i}^{\gamma}\tau_{j}^{\gamma}\right)\right.\nonumber \\
&+& \left.\left(r_{3}+r_{4}\right)\Pi_{s}^{(ij)}
\left(\frac{1}{2}-\tau_{i}^{\gamma}\right)
\left(\frac{1}{2}-\tau_{j}^{\gamma}\right)\right\} + {\cal H}_0\,,  
\end{eqnarray}
Each bond $\langle ij\rangle$ connects NN sites $\{i,j\}$ along one 
of the orthogonal axes $\gamma=a,b$ in the $ab$ plane. 
The model describes the spin-orbital superexchange in K$_2$CuF$_4$ 
\cite{Mos04}, with the superexchange constant $J=4t^2/U$.
The coefficients $r_1\equiv 1/(1-3\eta)$, $r_2=r_3\equiv 1/(1-\eta)$, 
$r_4\equiv 1/(1+\eta)$ refer to the 
$d_{i}^{9}d_{j}^{9}\rightleftharpoons d_{i}^{8}d_{j}^{10}$ charge 
excitations to the upper Hubbard band \cite{Ole00} and depend on Hund's 
exchange parameter 
\begin{equation}
\eta=\frac{J_{H}}{U}.
\label{eq:eta}
\end{equation}
The spin projection operators $\Pi_{ij}^{s}$ and $\Pi_{ij}^{t}$ select 
a singlet ($\Pi_{ij}^{s}$) or triplet ($\Pi_{ij}^{t}$) configuration
for spins $S=1/2$ on the bond $\langle ij\rangle$, respectively, 
\begin{equation}
\Pi_{s}^{(ij)}=\left(\frac{1}{4}-{\bf S}_{i}\cdot{\bf S}_{j}\right),\hskip.5cm
\Pi_{t}^{(ij)}=\left(\frac{3}{4}+{\bf S}_{i}\cdot{\bf S}_{j}\right).
\label{eq:proje}
\end{equation}
Here $\tau_{i}^{\gamma}$ act in the subspace of $e_{g}$ orbitals 
occupied by a hole $\{|x\rangle,|z\rangle\}$, with 
$|z\rangle\equiv(3z^2-r^2)/\sqrt{6}$ and 
$|x\rangle\equiv(x^2-y^2)/\sqrt{2}$ --- they can be expressed in terms of 
Pauli matrices $\{\sigma_{i}^{x},\sigma_{i}^{z}\}$ in the following way
\cite{Ole00}: 
\begin{equation}
\tau_{i}^{a(b)}\equiv\frac{1}{4}
\left(-\sigma_{i}^{z}\pm\sqrt{3}\sigma_{i}^{x}\right),
\quad\tau_{i}^{c}=\frac{1}{2}\,\sigma_{i}^{z}.
\label{eq:odefs}
\end{equation}
The term ${\cal H}_0$ in Eq. (1) is the crystal field splitting of two 
$e_g$ orbitals induced by the lattice geometry or pressure,
\begin{equation}
\label{eq:H0}
{\cal H}_0 = - E_{z}\sum_{i}\tau_{i}^{c}.
\end{equation}
When $|E_{z}|\gg J$ it dictates the FO order with either $z$ or $x$ 
orbitals as long as we stay in the AF regime. 
This ground state can be further improved using perturbation theory 
in a dimensionless parameter $|\varepsilon_z|^{-1}\equiv J/|E_{z}|$.

%%%%%%%%%%%%%%%%%%%%%%%%%%%%%%%%%%%%%%%%%%%%%%%%%%%%%%%%%%%%%%%%%%%%%%%%%
%                             figure 2
%%%%%%%%%%%%%%%%%%%%%%%%%%%%%%%%%%%%%%%%%%%%%%%%%%%%%%%%%%%%%%%%%%%%%%%%%
\begin{figure}[t!]
\includegraphics[width=8.4cm]{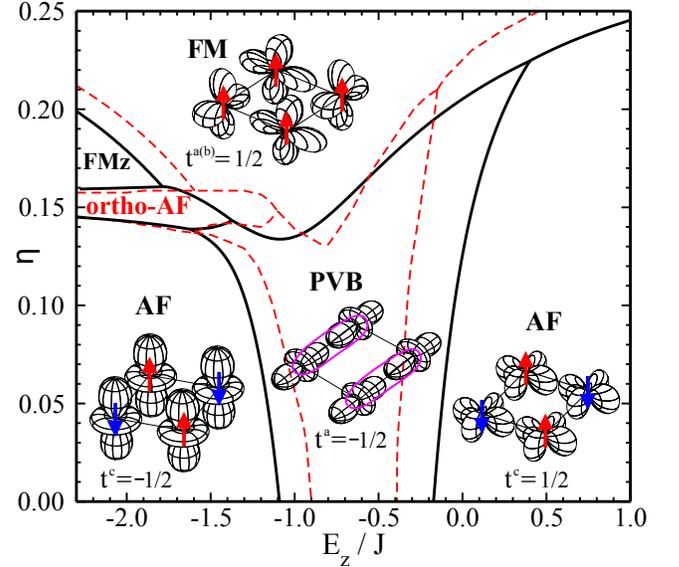}
\caption{(color online).
Phase diagram of the 2D KK model in the PMF (solid lines) and
ERA (dashed lines) variational approximation.
Insets show representative spin and orbital configurations on a 
$2\times2$ plaquette --- $x$-like ($t^c\!=\!\frac12$) and $z$-like 
($t^{a,c}\!=\!-\frac12$) 
orbitals \cite{t_vals} are accompanied either by AF spin order 
(arrows) or by spin singlets in the PVB phase (ovals). 
The FM phase has 
a two-sublattice AO order (with $t^{a(b)}\!=\!\frac12$ at $E_z=0$) 
or FO$z$ order (FM$z$). 
In between the AF and FM (FM$z$) phase on finds an exotic ortho-AF phase 
--- it has a noncollinear spin order, see text.
}
\label{fig:2ddiags_1} 
\end{figure}

In the PMF one finds self-consistently MFs:
$s_{i}^{\alpha}\equiv\langle S_{i}^{\alpha}\rangle$, 
$t_{i}^{\gamma}\equiv\langle\tau_{i}^{\gamma}\rangle$, and
$v_{i}^{\alpha,\gamma}\equiv\langle S_i^{\alpha}\tau_i^{\gamma}\rangle$.
Here $\gamma=a,b$, $\alpha=x,z$, and $i=1,\cdots,4$ labels sites of a 
single plaquette, see Fig. \ref{fig:ERA}(a). 
We assume that either all plaquettes are the same, or that 
the neighboring plaquettes are rotated by $\pi/2$ with respect to each 
other in the $ab$ plane. In the latter case the order parameters are 
interchanged ($a\leftrightarrow b$) between neighboring plaquettes and 
transform as: $\{t_{i}^{a(b)}\}\to\{t_{i}^{b(a)}\}$ and 
$\{v_{i}^{\alpha,a(b)}\}\to\{v_{i}^{\alpha,b(a)}\}$. 
In the ERA treatment we either assume that all 
${\cal P}$'s and ${\cal U}$'s are the same, like in Fig. \ref{fig:ERA}(b), 
or divide the plaquette 
lattices of ${\cal P}$'s and ${\cal U}$'s into four sublattices with four 
independent ${\cal P}$'s and ${\cal U}$'s. 
One finds that the energy found in the ERA, when optimized with respect 
to both ${\cal U}$ and ${\cal P}$, is typically $5...15\%$ lower than 
the one in the PMF. 

{\it Phase diagram.---}
The phase diagram in $(\varepsilon_z,\eta)$ plane contains six phases, 
see Fig. \ref{fig:2ddiags_1}. 
The same phases appear in both the PMF and ERA --- this suggests that 
the phase diagram is complete. At large $\eta$ one finds two FM phases: 
either with alternating orbital (AO) order as observed in K$_2$CuF$_4$ 
\cite{Ish96,Ish98} or with FO$z$ order (FM$z$). At $E_z<-1.8J$ 
a second order transition occurs from the FM to the FM$z$ phase
(all other transitions involve both spins and orbitals are first order) 
at $\varepsilon_z/r_1=-0.934$ $(0.837)$ in the PMF (ERA).
In these phases $\Pi^{(ij)}_t=1$ and the Hamiltonian (\ref{eq:hamik}) 
reduces to the $e_g$ orbital model \cite{You07} or 
to the generalized compass model \cite{Cin10}, in transverse field $E_z$. 
At $E_z=0$ one finds
AO order with $\langle\sigma^x\rangle\neq 0$, while finite $E_z$ induces 
transverse polarization $\langle\sigma^z\rangle\neq 0$. 

The phase diagram includes also two AF phases. They have uniform 
FO order with $\langle\sigma^z\rangle>0$ for $E_{z}>0$
and $\langle\sigma^z\rangle<0$ for $E_{z}<0$. 
The spin interactions in two AF phases are 
nonequivalent and are much weaker for $E_{z}<0$ than for $E_{z}>0$ --- 
this difference increases up to a factor of 9 for fully polarized 
orbitals \cite{Ez>0}. These two phases are separated by the plaquette 
valence bond (PVB) phase with pairs of parallel spin singlets, 
horizontal or vertical and alternating between NN plaquettes. Note 
that the PVB phase is an analog of spin liquid phases found before 
for the 3D KK model \cite{Fei97} and for the bilayer \cite{Brz11}.

%%%%%%%%%%%%%%%%%%%%%%%%%%%%%%%%%%%%%%%%%%%%%%%%%%%%%%%%%%%%%%%%%%%%%%%%%
%                             figure 3
%%%%%%%%%%%%%%%%%%%%%%%%%%%%%%%%%%%%%%%%%%%%%%%%%%%%%%%%%%%%%%%%%%%%%%%%%
\begin{figure}[t!]
\begin{center}\includegraphics[width=8.2cm]{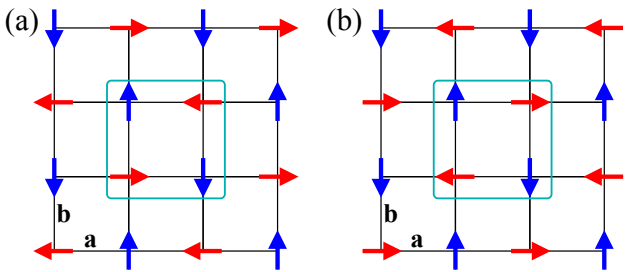}\end{center}
\caption{(color online). Schematic view of two nonequivalent spin
configurations (a) and (b) of the classical ortho-AF phase
$|\mathrm{AF}_{\perp}\rangle$ which cannot be
transformed one into the other by lattice translations. 
Four spin directions (arrows) correspond to four spin sublattices ---
up/down arrows stand for eigenstates of 
$\left\langle S_{i}^{z}\right\rangle =\pm\frac12$, 
while right/left arrows for 
$\left\langle S_{i}^{x}\right\rangle =\pm\frac12$. 
}
\label{fig:cls_daf} 
\end{figure}

Finally, at $E_{z}<0$ and $\eta\sim 0.15$ we find a novel exotic 
"orthogonal AF" (``ortho-AF'') phase 
with entanglement ($v_{i}^{\alpha,\gamma}\neq 0$)
which emerges in between the AF and the FM (FM$z$) 
phase. This state is characterized by the noncollinear magnetic 
order, see Fig. \ref{fig:cls_daf}, with NN spins being orthogonal 
to each other and next-nearest neighbor (NNN) spins being AF. This 
phase is robust and has a somewhat extended
range of stability in the 
ERA. In contrast to the frustrated spin $J_1$-$J_2$ 
interactions on a square lattice \cite{J1J2}, one finds here that $J_1$ 
is negligible and spin order follows from further neighbor couplings. 

{\it Effective spin model.---}
To explain the exotic magnetic order in the ortho-AF phase shown in Fig. 
\ref{fig:cls_daf} we derive an effective spin model for this phase. 
We show that NNN and third NN (3NN) spin interactions emerge here from 
the frustrated spin-orbital superexchange, 
${\cal V}\equiv {\cal H}-{\cal H}_{0}$, treated as perturbation of the
orbital ground state $|0\rangle$ of the unperturbed Hamiltonian 
${\cal H}_0$ (\ref{eq:H0}). Note an analogy to hidden multiple-spin 
interactions derived recently for frustrated Kondo lattice models 
\cite{Mot12}.

For negative $\varepsilon_z<0$ the ground state $|0\rangle$ of 
${\cal H}_0$ is the FO$z$ state with $z$ orbitals occupied by a hole at 
each site,  
$\tau_{i}^{c}\left|0\right\rangle=-\frac{1}{2}\left|0\right\rangle$, 
and the energy $\varepsilon_0=-\frac{1}{2}|\varepsilon_z|$ 
per site. A finite gap that 
occurs for orbital excitations helps to remove high spin degeneracy in 
$|0\rangle$ by effective spin interactions in the Hamiltonian $H_{s}$ 
that can be constructed using the expansion in powers of 
$|\varepsilon_z|^{-1}$,
\begin{equation}
\label{eq:Hes}
H_{s} \simeq J\left\{N\varepsilon_0+H_s^{(1)}+H_s^{(2)}+H_s^{(3)}\right\},
\end{equation}
where $N$ is the number of sites. The first order term is an average
$H_{s}^{(1)}\equiv \langle 0|{\cal V}|0\rangle$. Similarly, to evaluate 
$H_{s}^{(2)}$ and $H_{s}^{(3)}$ we determine the matrix 
elements $\langle n|{\cal V}|0\rangle$ for the excited states 
$|n\rangle$ with certain number of $z$-orbitals flipped to $x$-orbitals. 
All the averages are taken between orbital states and the spin model 
Eq. (\ref{eq:Hes}) follows.

The first order yields the Heisenberg Hamiltonian
\begin{equation}
\label{Hs1}
H_{s}^{(1)}=\frac{1}{2^{5}}\left(-3r_{1}+4r_{2}+r_{4}\right)
\sum_{\left\langle ij\right\rangle }
\left({\bf S}_{i}\cdot{\bf S}_{j}\right).
\end{equation}
The NN interaction $J_1\equiv(-3r_1+4r_2+r_4)J/2^5$ changes sign at 
$\eta_{0}\simeq 0.155$ implying a direct AF-FM transition. However, 
this turns out to be a premature conclusion because the vanishing of 
$H_s^{(1)}$ at $\eta_0$ makes higher order terms in Eq. (\ref{eq:Hes}) 
relevant. Indeed, $\eta_{0}$ nicely falls into the ortho-AF area of the 
phase diagram in Fig. \ref{fig:2ddiags_1}, where the NN interaction 
$J_1$ is small and frustrated.   

%%%%%%%%%%%%%%%%%%%%%%%%%%%%%%%%%%%%%%%%%%%%%%%%%%%%%%%%%%%%%%%%%%%%%%%%%
%                             figure 4
%%%%%%%%%%%%%%%%%%%%%%%%%%%%%%%%%%%%%%%%%%%%%%%%%%%%%%%%%%%%%%%%%%%%%%%%%
\label{sub:chir2d}
\begin{figure}[t!]
\begin{center}\includegraphics[width=8.4cm]{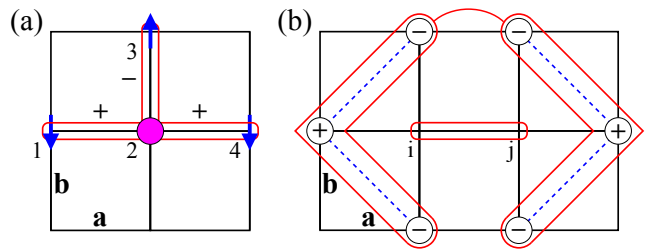}\end{center} 
\caption{(color online). 
Artist's views of the effective spin interactions obtained in: 
(a) second order $H_s^{(2)}$, with NNN term (123) and 3NN term (124), and
(b) third order $H_s^{(3)}$. 
The frames in (a) indicate Heisenberg bonds multiplied along $a$ or $b$ 
axes with $\pm$ sign depending on the bond direction; the dot in the 
center stands for an orbital flip in $\left|0\right\rangle$. 
In (b) the dashed lines symbolize sums of three spins which enter each 
effective spin, ${\bf S}_{{\cal N}_{\gamma}(i)}$ and 
${\bf S}_{{\cal N}_{\bar{\gamma}}(j)}$; the phase factors $s_{\gamma}$ 
(circles) and their scalar product are marked with connected frames.
}
\label{fig:pert2} 
\end{figure}

{\it Higher order terms.---}
Higher order terms arise by flipping orbitals from the ground state 
$|0\rangle$. Given that ${\cal V}$ has non-zero overlap only with 
states having one or two NN orbitals flipped from $z$ to $x$, 
one finds in second order,
\begin{equation}
\label{Hs2}
H_{s}^{(2)}=\frac{\xi(\eta)}{|\varepsilon_z|}\left\{ 
\sum_{\left\langle \left\langle ij\right\rangle \right\rangle }\!
\left({\bf S}_{i}\cdot{\bf S}_{j}\right)-\tfrac{1}{2}\!
\sum_{\left\langle \left\langle \left\langle ij
     \right\rangle \right\rangle \right\rangle }
\!\left({\bf S}_{i}\cdot{\bf S}_{j}\right)\right\} ,
\end{equation}
with 
$\xi(\eta)=(r_{1}+2r_{2}+3r_{4}){}^2/2^{10}$.
Here $\langle\langle ij\rangle\rangle$ and 
$\langle\langle\langle ij\rangle\rangle\rangle $
stand for NNN and 3NN sites $i$ and $j$, see Fig. 
\ref{fig:pert2}(a) for the origin and sign of these interactions. Apart 
from this, the second order also brings the $|\varepsilon_{z}|^{-1}$ 
correction to the Heisenberg interactions of $H_{s}^{(1)}$ (\ref{Hs1}), 
moving the transition point from $\eta_{0}$ to 
$\eta_{0}+O\left(\varepsilon_{z}^{-1}\right)$. 

The NNN AF interaction in $H_s^{(2)}$ (\ref{Hs2}) alone would give two 
quantum antiferromagnets on interpenetrating sublattices \cite{spiral},
but the additional 3NN FM term makes these AF states more classical 
than in the 2D Heisenberg model \cite{Suppl}. This ``double-AF'' 
configuration is already similar to the ortho-AF phase in Fig. 
\ref{fig:cls_daf}. However, the second order does not explain why the 
spins in the ortho-AF phase prefer to be orthogonal on NN bonds,
and we have to proceed to the third order.

The third order in Eq. (\ref{eq:Hes}) produces many contributions to 
the spin Hamiltonian, but we are interested only in 
qualitatively new terms comparing to the lower orders. The terms bringing 
potentially new physics are the ones with
connected products of three different Heisenberg bonds \cite{Suppl}.
The final result is a four-spin coupling,
\begin{equation}
H{}_{\perp}^{(3)} = \frac{1}{\varepsilon_{z}^{2}}\,\chi(\eta)\xi(\eta)
\sum_{\langle ij\rangle||\gamma}\left({\bf S}_{i}\cdot{\bf S}_{j}\right) 
\left({\bf S}_{{\cal N}_{\gamma}(i)}\cdot{\bf S}_{{\cal N}_{\bar{\gamma}}(j)}\right),
\label{eq:resultA-1}
\end{equation}
where $\bar{\gamma}=-\gamma$ and 
${\bf S}_{{\cal N}_{\gamma}(i)}\equiv\sum_{\alpha\not=\gamma}
s_{\alpha}{\bf S}_{i+\alpha}$ is an effective spin
around site $i$ in the direction $\gamma$, see Fig. \ref{fig:pert2}(b).
Here $\chi(\eta)=9(r_{1}+r_{4})/2^7$, 
$\alpha\in\{\pm a,\pm b\}$, and $s_{\alpha}=-1$ for 
$\alpha=\pm b$ and $s_{\alpha}=1$ otherwise.
In the limit of two interpenetrating classical antiferromagnets 
$H_{\perp}^{(3)}$ gives the energy per site,
$\varepsilon_{\perp}^{(3)}\approx\varepsilon_{z}^{-2}
\chi(\eta)\xi(\eta)(\frac{3}{4}\cos\varphi)^2$,
where $\varphi$ is an angle between the NN spins \cite{Suppl}. This 
classical energy is minimized for $\varphi=\pi/2$ which explains the 
exotic magnetic order in the ortho-AF phase, shown in Fig. 
\ref{fig:cls_daf}.

%%%%%%%%%%%%%%%%%%%%%%%%%%%%%%%%%%%%%%%%%%%%%%%%%%%%%%%%%%%%%%%%%%%%%%%%%
%                             figure 5
%%%%%%%%%%%%%%%%%%%%%%%%%%%%%%%%%%%%%%%%%%%%%%%%%%%%%%%%%%%%%%%%%%%%%%%%%
\begin{figure}[t!]
\includegraphics[width=6.3cm]{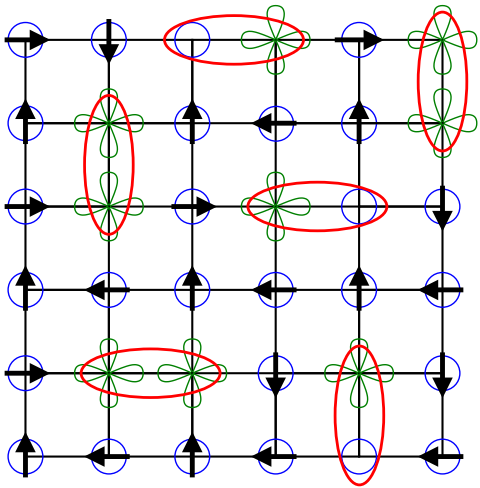} 
\caption{(color online).
Artist's view of the ortho-AF state $\left|\Psi_{\rm SO}\right\rangle$
(\ref{so}), with spin order (arrows) of Fig. \ref{fig:cls_daf} and FO$z$ 
orbital order (circles) in the $ab$ plane. The state is dressed with 
spin singlets (ovals) entangled with either one or two orbital 
excitations from $|z\rangle$ to $|x\rangle$ orbitals (clovers) on NN 
bonds.
}
\label{fig:sing_orb} 
\end{figure}

{\it Spin-orbital entanglement.---}
The ground state $\left|\mathrm{AF}_{\perp}\right\rangle$ of $H_s$ 
(\ref{eq:Hes}) is nearly classical, except for small 
quantum corrections obtained within the spin-wave expansion 
\cite{Suppl}. Thus one might expect that the spins are 
not entangled with orbitals. However, this argument overlooks that the 
resulting spins in $H_s$ are dressed with orbital and spin-orbital 
fluctuations. Indeed, within the perturbative treatment we obtain the 
full spin-orbital ground state, 
\begin{equation}
\label{so}
\left|\Psi_{\rm SO}\right\rangle\propto 
\left(1-
\sum_{n\not=0}   \frac{{\cal V}_n}{\varepsilon_{n}}+
\sum_{n,m\not=0} \frac{{\cal V}_n{\cal V}_m}{\varepsilon_{n}\varepsilon_{m}}
-\dots\right)\left|\Phi_0\right\rangle,
\end{equation}
where ${\cal V}_n\equiv\left|n\right\rangle\left\langle n\right|{\cal V}$, 
$\varepsilon_n$ are excitation energies, and 
$\left|\Phi_0\right\rangle\equiv|\mathrm{AF}_{\perp}\rangle|0\rangle$
is the disentangled classical state (Fig. \ref{fig:cls_daf}).
The operator sum in front of $\left|\Phi_0\right\rangle$ dresses this 
state with both orbital and spin-orbital fluctuations. When the purely
orbital fluctuations are neglected and density of spin-orbital defects 
is assumed to be small, one finds 
\begin{equation}
\label{eq:ldAp}
\left|\Psi_{\rm SO}\right\rangle\simeq\,
\exp\left( -\frac{1}{|\varepsilon_z|}
\sum_{\langle ij\rangle||\gamma}{\cal D}_{ij}^{\gamma}\right) 
\left|\Phi_0\right\rangle,
\end{equation}
where
\begin{equation}
\label{Dij}
{\cal D}_{ij}^{\gamma}=\left\{
-A\,\sigma_{i}^{x}\sigma_{j}^{x}
+B\left(\sigma_{i}^{x}+\sigma_{j}^{x}\right)s_{\gamma}\right\}\Pi_{ij}^{s}
\end{equation}
is the spin-orbital excitation operator on the bond $\langle ij\rangle$, with
$A=3(r_{1}+r_{4})/2^6$ and $B=\sqrt{3}(r_{1}+2r_{2}+3r_{4})/2^5$. 
Both terms in Eq. (\ref{Dij}) project on a NN spin singlet, but the first 
one flips two NN $z$-orbitals while the second one generates 
only one flipped orbital. In short, the 
exponent $e^{-D/|\varepsilon_z|}$ dresses the classical ortho-AF state 
$|{\rm AF}_{\perp}\rangle$ in Fig. \ref{fig:cls_daf} with the entangled 
(spin-singlet/flipped-orbital) defects, see
Fig. \ref{fig:sing_orb}. The density of such entangled defects increases 
when $|\varepsilon_z|$ is decreased towards the PVB phase. 

{\it Topological defects.---}
The order parameter of the ortho-AF phase has non-trivial topology. 
The ground state is degenerate with respect to different orientations 
of its order parameter that consists of two orthogonal unit vectors 
defining orientation of each antiferromagnet. The first vector 
lives on the whole sphere $S^2$, but the second one is restricted to 
a circle $S^1$ because it is orthogonal to the first. In addition to 
spin-wave excitations, this $S^2\times S^1$ topology allows for 
skyrmions (textures) \cite{Wil83} and $\mathbb{Z}_2$-vortices 
(hedgehogs) as two types of topological defects. The hedgehog is 
stabilized by the orthogonality of the antiferromagnets. For instance, 
when one of them has fixed uniform orientation of its N\'eel order in 
space, the orthogonal orientation of the other one is free to make a 
hedgehog-like rotation.

{\it Summary.---}
We have found surprising {\it noncollinear\/} spin order that 
arises from the NN spin-orbital superexchange when ferromagnetic and 
antiferromagnetic interactions almost compensate each other in the 
2D KK model away from orbital degeneracy. It is stabilized by further 
neighbor spin exchange generated by entangled spin-orbital fluctuations 
which involve spin singlets and orbital flips. Similar mechanism works 
in the 3D KK model where it leads to a rich variety of spin-orbital 
phases to be reported elsewhere. 

Finally, we note that magnetic order in spin-orbital systems may be 
changed by applying pressure \cite{Goode} --- indeed a transition from 
ferromagnetic to antiferromagnetic order was observed in K$_2$CuF$_4$ 
\cite{Ish96,Ish98}. Such a transition is also found here for a 
realistic value of $\eta\simeq 0.15$, and one could induce it in the 
antiferromagnetic phase by external magnetic field.
Whether the antiferromagnetic order could be 
noncollinear as predicted here remains an experimental challenge.

{\it Acknowledgments.---}
We thank G. Khaliullin, R. Kremer and B. Normand 
for insightful discussions.
This work was supported by the Polish National Science Center 
(NCN) under Projects No. N202 069639 (W.B. and A.M.O.) and
2011/01/B/ST3/00512 (J.D.).

\newpage
                   
\section*{Supplemental Material }

%\author{Wojciech Brzezicki}

%\author{Jacek Dziarmaga}

%\author{Andrzej M. Ole\'{s} }

This supplement presents the technical details of the analysis 
employed in the paper. We first consider the third order terms in 
the perturbative expansion in section A. They justify the angle 
$\varphi=\pi/2$ obtained for the nearest neighbor (NN) spins in the 
regime of the noncollinear orthogonal antiferromagnetic (ortho-AF) 
phase. In section B we develop the spin-wave theory for the ortho-AF 
phase and calculate the quantum corrections to the order parameter. 
These calculations show that the noncollinear ortho-AF phase is stable 
with respect to the Gaussian fluctuations and the quantum corrections 
are here weaker than for the two-dimensional (2D) AF Heisenberg model.

\subsection{ Third order terms in the perturbative expansion }
\label{sec:pert}

\begin{figure}[t!]
\begin{centering}
\includegraphics[width=8cm]{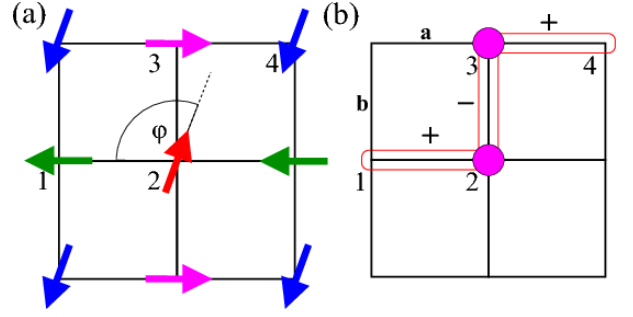}
\par\end{centering}
\caption{
Panel (a): two independent AF orders realized by effective 
spin Hamiltonian $H_{s}$ up to second order. Angle $\varphi$ between 
the two neighboring spins is undetermined. 
Panel (b): exemplary third order correction
to $H_{s}$ fixing the angle $\varphi$ as $\varphi=\pi/2$. Red frames
stand for Heisenberg bond with $\pm$ sign depending on the bond's
direction and magenta dots indicate the single-site orbital 
excitations in the ground state.}
\label{fig:chirang} 
\end{figure}

The second order in the perturbation theory in $|\varepsilon_z|^{-1}$ 
results in two antiferromagnets on interpenetrating sublattices, but 
the angle $\varphi$ between the nearest neighbor (NN) spins
remains undetermined, see Fig.
\ref{fig:chirang}(a). Thus we have to consider third order 
contributions to the effective spin Hamiltonian $H_{s}$ of the form: 
\begin{eqnarray}
H^{(3)}_s\! & = &\! \frac{1}{\varepsilon_z^2}\sum_{i,\gamma'}s_{\gamma'}
\left[-J^{xz}\!(\eta){\bf S}_{i}\cdot{\bf S}_{i+\gamma'}
+K^{xz}\!(\eta)\right]\nonumber \\
 & \times &\! \sum_{\gamma\not=\gamma'}\left[J^{xx}\!(\eta){\bf S}_{i}
 \cdot{\bf S}_{i+\gamma}\!+K^{xx}\!(\eta)\right]\nonumber \\
 & \times &\! \sum_{\gamma''\not=-\gamma}
 s_{\gamma''}\left[-J^{xz}\!(\eta){\bf S}_{i+\gamma}\cdot
 {\bf S}_{i+\gamma+\gamma''}\!+K^{xz}\!(\eta)\right]\nonumber \\
 & + &\! \frac{1}{2\varepsilon_z^2}\sum_{i,\gamma}\left[J^{xx}(\eta)
 {\bf S}_{i}\cdot{\bf S}_{i+\gamma}\!+K^{xx}\!(\eta)\right]\nonumber \\
 & \times &\! \sum_{\gamma'\not=\gamma}s_{\gamma'}
 \left[-J^{xz}(\eta){\bf S}_{i}\cdot{\bf S}_{i+\gamma'}
 +K^{xz}\!(\eta)\right]\nonumber \\
 & \times &\!\! \sum_{\gamma''\not=-\gamma}\!
 s_{\gamma''}\left[-J^{xz}\!(\eta){\bf S}_{i+\gamma}\cdot
 {\bf S}_{i+\gamma+\gamma''}\!+K^{xz}\!(\eta)\right],
\end{eqnarray}
where: $J^{xz}(\eta)=2^{-5}\sqrt{3}\left(r_{1}\!+2r_{2}\!+3r_{4}\right)$,
$K^{xz}(\eta)=2^{-7}\sqrt{3}\left(2r_{2}\!-3r_{1}+\!3r_{4}\right)$,
$J^{xx}(\eta)=2^{-5}3\left(r_{1}+r_{4}\right)$ and 
$K^{xx}(\eta)=2^{-7}3\left(3r_{1}-r_{4}\right)$.
Here and in all other equations in this Section a sum over $\gamma$ 
means the sum over all directions in the square lattice, i.e., 
$\gamma=\pm a,\pm b$.

The spin chains with less than three scalar products do not contribute
with any qualitatively new terms. Once they are omitted we obtain: 
\begin{eqnarray}
H_{{\rm chain}}^{(3)}\! & = & \!
\frac{9}{2^{16}\varepsilon_z^2}\left(r_{1}\!+\! r_{4}\right)
\left(r_{1}\!+\!2r_{2}\!+\!3r_{4}\right)^{2}\sum_{i,\gamma}
\sum_{{\gamma'\not=\gamma\atop \gamma''\not=-\gamma}}\nonumber \\
&  & \!\left\{2s_{\gamma}s_{\gamma''}
 \left({\bf S}_{i+\gamma}\cdot{\bf S}_{i}\right)
 \left({\bf S}_{i}\cdot{\bf S}_{i+\gamma'}\right)\!\left({\bf S}_{i+\gamma'}\cdot
 {\bf S}_{i+\gamma'+\gamma''}\right)\right.\nonumber \\
 & + & \!s_{\gamma'}s_{\gamma''}\left.\left(
 {\bf S}_{i}\cdot{\bf S}_{i+\gamma}\right)\!\left({\bf S}_{i+\gamma'}\cdot
 {\bf S}_{i}\right)\!\left({\bf S}_{i+\gamma}\cdot{\bf S}_{i+\gamma+\gamma''}
\right)\right\},\nonumber \\
\label{eq:3tild}
\end{eqnarray}
where $s_{\gamma}$ is a sign factor depending on bond's direction
$\gamma$ originating from the definition of operators $\tau_{i}^{a,b}$
[see Fig. \ref{fig:chirang}(b)], i.e., 
\begin{equation}
s_{\gamma}=\left\{ \begin{array}{ccc}
1 & {\rm if} & \gamma=\pm a\\
-1 & {\rm if} & \gamma=\pm b
\end{array}\right..
\end{equation}
We transform the second term of Eq. (\ref{eq:3tild}) using the vector 
identity:
\begin{eqnarray}
\left({\bf S}_{i}\cdot{\bf S}_{i+\gamma}\right)
 \left({\bf S}_{i+\gamma'}\cdot{\bf S}_{i}\right)&=&\nonumber \\
\left({\bf S}_{i+\gamma'}\cdot{\bf S}_{i}\right)
 \left({\bf S}_{i}\cdot{\bf S}_{i+\gamma}\right)&+&
i{\bf S}_{i+\gamma}({\bf S}_{i+\gamma'}\times{\bf S}_{i}).
\end{eqnarray}
The antihermitian term with a cross product cancels out under the
sum in Eq. (\ref{eq:3tild}), thus we obtain: 
\begin{eqnarray}
H{}_{{\rm chain}}^{(3)}\! & = & \!\frac{9}{2^{16}\varepsilon_z^2}
\left(r_{1}\!+\! r_{4}\right)\left(r_{1}\!+\!2r_{2}\!+\!3r_{4}\right)^{2}
\sum_{i,\gamma}\sum_{{\gamma'\not=\gamma\atop \gamma''\not=-\gamma}}\nonumber \\
&  & \!\left\{2s_{\gamma}s_{\gamma''}\!\left(
 {\bf S}_{i+\gamma}\cdot{\bf S}_{i}\right)\!\left({\bf S}_{i}\cdot{\bf S}_{i+\gamma'}\right)
 \left({\bf S}_{i+\gamma'}\cdot{\bf S}_{i+\gamma'+\gamma''}\right)\right.\nonumber \\
& + & \! s_{\gamma'}s_{\gamma''}\left.\left({\bf S}_{i+\gamma'}\cdot{\bf S}_{i}\right)
 \left({\bf S}_{i}\cdot{\bf S}_{i+\gamma}\right)
 \left({\bf S}_{i+\gamma}\cdot{\bf S}_{i+\gamma+\gamma''}\right)\right\}.\nonumber \\
\label{eq:3tild2}
\end{eqnarray}
Now all the scalar products are ordered along the lines: 
$\left(i+\gamma\right)\to\left(i\right)\to
\left(i+\gamma'\right)\to\left(i+\gamma'+\gamma''\right)$
and $\left(i+\gamma'\right)\to\left(i\right)\to\left(i+\gamma\right)
\to\left(i+\gamma+\gamma''\right)$.
Next we use another spin identity, namely 
\begin{eqnarray}
&  & \left({\bf S}_{1}\cdot{\bf S}_{2}\right)\left(
{\bf S}_{2}\cdot{\bf S}_{3}\right)\left({\bf S}_{3}\cdot{\bf S}_{4}\right)=\nonumber \\
&  & \frac{1}{16}\left({\bf S}_{1}\cdot{\bf S}_{4}\right)
+\frac{1}{4}\left({\bf S}_{1}\cdot{\bf S}_{4}\right)\left(
{\bf S}_{2}\cdot{\bf S}_{3}\right)-\frac{1}{4}\left({\bf S}_{1}\cdot{\bf S}_{3}\right)
\left({\bf S}_{2}\cdot{\bf S}_{4}\right)\nonumber \\
&  & +\frac{i}{8}\,{\bf S}_{1}\left({\bf S}_{3}\times{\bf S}_{4}\right)
+\frac{i}{8}\,{\bf S}_{1}\left({\bf S}_{4}\times{\bf S}_{2}\right).
\label{eq:3sp_id}
\end{eqnarray}
Again, the antihermitian cross-product terms cancel out under the
sums in $H{}_{{\rm chain}}^{(3)}$.

To analyze the relevance of other terms in Eq. (\ref{eq:3sp_id})
we have to take into account that, to second order in the perturbation
theory, there is nearly classical AF order on the two sublattices.
We observe that:\\ 
(i) the first term is an AF interaction between the sublattices which 
\textit{is not compatible} with the antiferromagnetism on the 
sublattices that is ${\cal O}(E_{z})$ stronger, \\
(ii) depending
on its sign the second term may favour orthogonality of the two AF
orders which \textit{is compatible} with the order on sublattices, and \\
(iii) the third term brings no new information about the order.\\
Taking into account all three above arguments we argue that the relevant
type-(ii) third order perturbative contributions of the form given by 
Eq. (\ref{eq:3sp_id}) may favour orthogonality of the two AF orders.
Now we have to extract all such contributions from Eq. (\ref{eq:3sp_id})
and check if their overall sign is indeed positive.

After transforming Eq. (\ref{eq:3tild2}) we obtain 
\begin{eqnarray}
H_{\perp}^{(3)} & = & \frac{9}{2^{18}\varepsilon_z^2}
\left(r_{1}\!+\! r_{4}\right)\left(r_{1}\!+\!2r_{2}\!+\!4r_{4}\right)^{2}
\sum_{i,\gamma}\sum_{{\gamma'\not=\gamma\atop \gamma''\not=-\gamma}}\nonumber \\
&  & \left[2s_{\gamma}s_{\gamma''}\!\left({\bf S}_{i+\gamma}\cdot
{\bf S}_{i+\gamma'+\gamma''}\right)
\left({\bf S}_{i}\cdot{\bf S}_{i+\gamma'}\right)\right.\nonumber \\
& + &  s_{\gamma'}s_{\gamma''}\left.\left({\bf S}_{i+\gamma'}\cdot
{\bf S}_{i+\gamma+\gamma''}\right)\!\left({\bf S}_{i}\cdot
{\bf S}_{i+\gamma}\right)\right],
\label{eq:3perp}
\end{eqnarray}
or in a more compact form 
\begin{eqnarray}
H_{\perp}^{(3)} & = & \frac{27}{2^{18}\varepsilon_z^2}
\left(r_{1}\!+\! r_{4}\right)\left(r_{1}\!+\!2r_{2}\!
+\!4r_{4}\right)^{2}\sum_{i,\gamma}\left({\bf S}_{i\cdot}
{\bf S}_{i+\gamma}\right)\nonumber \\
 & \times & \left(\sum_{\gamma'\not=\gamma}s_{\gamma'}
 {\bf S}_{i+\gamma'}\right)\!\!\cdot\!\!\left(\sum_{\gamma''\not=-\gamma}
 s_{\gamma''}{\bf S}_{i+\gamma+\gamma''}\right).
 \label{eq:resultA}
\end{eqnarray}
For two interpenetrating classical antiferromagnets $H_{\perp}^{(3)}$
gives the energy per site,
\begin{equation}
\varepsilon_{\perp}^{(3)}\approx\frac{1}{\varepsilon_z^2}\chi(\eta)
\xi(\eta)\left(\frac{3}{4}\cos\varphi\right)^2~.
\end{equation}
This classical energy is minimized for $\varphi=\pi/2$, i.e., when the
NN spins on the bonds are orhogonal. This completes the argument that 
the orders in the two antiferromagnets prefer to be orthogonal.

\subsection{Spin wave expansion in the noncollinear ortho-AF phase}
\label{sec:sw}

We start from the general form of the effective spin Hamiltonian:
\begin{eqnarray}
H_{s} & = &\! A\sum_{\langle ij\rangle}{\bf S}_{i}\cdot{\bf S}_{j}+2B\!
\sum_{\langle\langle ij\rangle\rangle }\!
{\bf S}_{i}\cdot{\bf S}_{j}-B\!\sum_{
\langle\langle\langle ij\rangle\rangle\rangle }\!
{\bf S}_{i}\cdot{\bf S}_{j}\label{eq:HS} \nonumber \\
& + &\! C\sum_{\langle ij\rangle||\gamma}\left({\bf S}_{i}\cdot
{\bf S}_{j}\right)\!\!\sum_{{\gamma'\not=\gamma\atop \gamma''\not=-\gamma}}\!
s_{\gamma'}s_{\gamma''}\left({\bf S}_{i+\gamma'}\cdot
{\bf S}_{i+\gamma+\gamma''}\right),\nonumber \\
\end{eqnarray}
with coeffcients $A$, $B$ and $C$ being the functions of $\eta$ and 
$\varepsilon_z$, i.e.,
\begin{eqnarray}
A & = & \frac{1}{2^{5}}\left(-3r_{1}+4r_{2}+r_{4}\right)\nonumber\\
 & + & \frac{1}{\varepsilon_z}\frac{3}{2^{11}}
\left\{3\left(r_{1}^{2}-r_{4}^{2}\right)-2
\left(r_{1}+2r_{2}+3r_{4}\right)^{2}\right\}, \\
B & = & -\frac{1}{\varepsilon_z}\frac{3}{2^{11}}\!
\left(r_{1}+2r_{2}\!+3r_{4}\right){}^{\!2}, \\
C & = & \frac{27}{2^{17}\varepsilon_z^{2}}\left(r_{1}
+ r_{4}\right)\left(r_{1}+2r_{2}+3r_{4}\right)^{2}.
\end{eqnarray}
To describe the ortho-AF order we divide the lattice into four 
sublattices as follows 
\begin{equation}
{\bf S}_{i,j}^{pq}\equiv{\bf S}_{p+2i,q+2j},\quad p,q\in\left\{ 1,2\right\} ,
\end{equation}
where $p$ and $q$ form the sublattice label. In what follows all
the sums over $p$ and $q$ run over the set $\{1,2\}$. Now, in each
sublattice we do the linearized Holstein-Primakoff transformation
around the ortho-AF order, i.e.,
\begin{equation}
\begin{array}{cccccc}
S_{i,j}^{x,11}\!\! & \mathbf{=}\!\! & \frac{1}{2}\!\!
\left(a_{i,j}^{11\dagger}+a_{i,j}^{11}\right), 
& S_{i,j}^{x,22} & \!\!\mathbf{=}\!\! & \frac{1}{2}
\left(a_{i,j}^{22\dagger}+a_{i,j}^{22}\right),\\
S_{i,j}^{y,11}\!\! & \mathbf{=}\!\! & 
\frac{1}{2i}\!\!\left(a_{i,j}^{11}-a_{i,j}^{11\dagger}\right), 
& S_{i,j}^{y,22} & \!\!\mathbf{=}\!\! & \frac{1}{2i}
\left(a_{i,j}^{22\dagger}-a_{i,j}^{22}\right),\\
S_{i,j}^{z,11}\!\! & \mathbf{=}\!\! & 
\left(\frac{1}{2}-a_{i,j}^{11\dagger}a_{i,j}^{11}\right), 
& S_{i,j}^{z,22} & \!\!\mathbf{=}\!\! & 
\left(a_{i,j}^{22\dagger}a_{i,j}^{22}-\frac{1}{2}\right),
\end{array}
\end{equation}
and
\begin{equation}
\begin{array}{cccccc}
S_{i,j}^{x,12}\!\! & \mathbf{=} & \!\!\left(a_{i,j}^{12\dagger}a_{i,j}^{12}
-\frac{1}{2}\right), & S_{i,j}^{x,21} & \!\!\mathbf{=}\!\! & 
\left(\frac{1}{2}-a_{i,j}^{21\dagger}a_{i,j}^{21}\right),\\
S_{i,j}^{y,12}\!\! & \mathbf{=} & \!\!\frac{1}{2}\!\!
\left(a_{i,j}^{12\dagger}+a_{i,j}^{12}\right), & S_{i,j}^{y,21} & 
\!\!\mathbf{=}\!\! & \frac{1}{2}\!\!
\left(a_{i,j}^{21\dagger}+a_{i,j}^{21}\right),\\
S_{i,j}^{z,12}\!\! & \mathbf{=} & \!\!\frac{1}{2i}\!\!
\left(a_{i,j}^{12\dagger}-a_{i,j}^{12}\right), & S_{i,j}^{z,21} 
& \!\!\mathbf{=}\!\! & \frac{1}{2i}\!\!\left(a_{i,j}^{21}-a_{i,j}^{21\dagger}\right).
\end{array}
\end{equation}
Next step is the Fourier transform (FT): 
\begin{equation}
a_{ij}^{pq}=\sum_{{\bf k}}e^{i(ik_{b}+jk_{a})}a_{{\bf k}}^{pq},
\end{equation}
followed by the phase transformation, in order to get rid of imaginary
parts in $H_s$ of Eq. (\ref{eq:HS}) after FT,
\begin{eqnarray}
a_{{\bf k}}^{11\dagger}\to a_{{\bf k}}^{11\dagger}e^{-i\frac{k_{a}}{2}}, 
& \qquad & a_{{\bf k}}^{12}\to ia_{k}^{12},\nonumber \\
a_{{\bf k}}^{21}\to-ia_{{\bf k}}^{21}e^{i\frac{k_{a}}{2}}e^{-i\frac{k_{b}}{2}}, 
& \qquad & a_{{\bf k}}^{22\dagger}\to a_{{\bf k}}^{22\dagger}e^{i\frac{k_{b}}{2}}. 
\end{eqnarray}
Finally, the interactions in the $H_s$ Hamiltonian take the following form:
\begin{eqnarray}
\sum_{\left\langle ij\right\rangle }\!{\bf S}_{i}\cdot{\bf S}_{j}\! & = & 
\!\frac{1}{2}\!\sum_{{\bf k}}\cos\frac{k{}_{b}}{2}\left(a_{-{\bf k}}^{21\dagger}\!
- a_{{\bf k}}^{21}\right)\left(a_{{\bf k}}^{11\dagger}\!- a_{-{\bf k}}^{11}\right)\nonumber \\
 & + & \!\frac{1}{2}\!\sum_{{\bf k}}\cos\frac{k{}_{b}}{2}\!\!
 \left(a_{-{\bf k}}^{12\dagger}\!- a_{{\bf k}}^{12}\right)\left(a_{{\bf k}}^{22\dagger}\!
 - a_{-{\bf k}}^{22}\right)\nonumber \\
 & - & \!\frac{1}{2}\!\sum_{{\bf k}}\cos\frac{k{}_{a}}{2}\!\!
 \left(a_{-{\bf k}}^{12\dagger}\!- a_{{\bf k}}^{12}\right)\left(a_{{\bf k}}^{11\dagger}\!
 - a_{-{\bf k}}^{11}\right)\nonumber \\
 & - & \!\frac{1}{2}\!\sum_{{\bf k}}\cos\frac{k{}_{a}}{2}\!
 \left(a_{-{\bf k}}^{21\dagger}\!- a_{{\bf k}}^{21}\right)\left(a_{{\bf k}}^{22\dagger}\!
 - a_{-{\bf k}}^{22}\right),\nonumber \\
\end{eqnarray}
for the NN interactions, 
\begin{eqnarray}
\sum_{\langle\langle ij\rangle\rangle }\!
{\bf S}_{i}\cdot{\bf S}_{j}\! & = & 2\sum_{{\bf k}}\left\{\gamma'_{{\bf k}}
\left(a_{{\bf k}}^{11\dagger}a{}_{-{\bf k}}^{22\dagger}
+ a_{{\bf k}}^{12\dagger}a_{-{\bf k}}^{21\dagger}\right)+H.c.\right\} \nonumber \\
\! & + & 2\sum_{{\bf k}}\sum_{p,q}a_{{\bf k}}^{pq\dagger}a_{{\bf k}}^{pq},
\end{eqnarray}
for the NNN interactions,
\begin{equation}
\sum_{\left\langle \left\langle \left\langle i,j\right\rangle \right\rangle \right\rangle }
{\bf S}_{i}\cdot{\bf S}_{j}=2\sum_{{\bf k}}
\sum_{p,q}a_{{\bf k}}^{pq\dagger}a_{{\bf k}}^{pq}\left(\gamma_{{\bf k}}-1\right)
\end{equation}
for the 3NN interactions, and
\begin{eqnarray}
& &\sum_{i,\gamma}\left({\bf S}_{i}\cdot{\bf S}_{i+\gamma}\right)s_{\gamma'}
s_{\gamma''}\sum_{{\gamma'\not=\gamma\atop \gamma''\not=-\gamma}}
\left({\bf S}_{i+\gamma'}\cdot{\bf S}_{i+\gamma+\gamma''}\right)  =\nonumber \\
& & 12\sum_{\bf k}\sum_{p,q}\gamma_{\bf k}\left(a_{\bf k}^{pq\dagger}+a_{-\bf k}^{pq}\right)
\left(a_{-\bf k}^{pq\dagger}+a_{\bf k}^{pq}\right) \nonumber \\
&-&48\sum_{\bf k}\gamma'_{k}\left(a_{\bf k}^{12\dagger}+a_{-\bf k}^{12}\right)
\left(a_{-\bf k}^{21\dagger}+a_{\bf k}^{21}\right) \nonumber \\
&-&48\sum_{\bf k}\gamma'_{\bf k}\left(a_{\bf k}^{11\dagger}+a_{-\bf k}^{11}\right)
\left(a_{-\bf k}^{22\dagger}+a_{\bf k}^{22}\right) \nonumber \\
&+&36\sum_{\bf k}\left(a_{\bf k}^{11\dagger}+a_{-\bf k}^{11}\right)
\left(a_{-\bf k}^{21\dagger}+a_{\bf k}^{21}\right)\cos\frac{k_{b}}{2} \nonumber \\
&+&36\sum_{\bf k}\left(a_{\bf k}^{12\dagger}+a_{-\bf k}^{12}\right)
\left(a_{-\bf k}^{22\dagger}+a_{\bf k}^{22}\right)\cos\frac{k_{b}}{2} \nonumber \\
&-&36\sum_{\bf k}\left(a_{\bf k}^{11\dagger}+a_{-\bf k}^{11}\right)
\left(a_{-\bf k}^{12\dagger}+a_{\bf k}^{12}\right)\cos\frac{k_{a}}{2} \nonumber \\
&-&36\sum_{\bf k}\left(a_{-\bf k}^{21\dagger}+a_{\bf k}^{21}\right)
\left(a_{\bf k}^{22\dagger}+a_{-\bf k}^{22}\right)\cos\frac{k_{a}}{2},
\end{eqnarray}
for the third order interactions between the two AF sublattices.
The coefficients $\{\gamma_{\bf k},\gamma'_{\bf k}\}$ are defined as: 
\begin{eqnarray}
\gamma_{\bf k} & \equiv & \frac{1}{2}\left(\cos k{}_{b}+\cos k{}_{a}\right),\\
\gamma'_{\bf k} & \equiv & \frac{1}{2}\left(\cos\frac{k_{a}+k_{b}}{2}
+\cos\frac{k_{a}-k_{b}}{2}\right).
\end{eqnarray} 

\begin{figure}[t!]
\begin{centering}
\begin{center}\includegraphics[width=7cm]{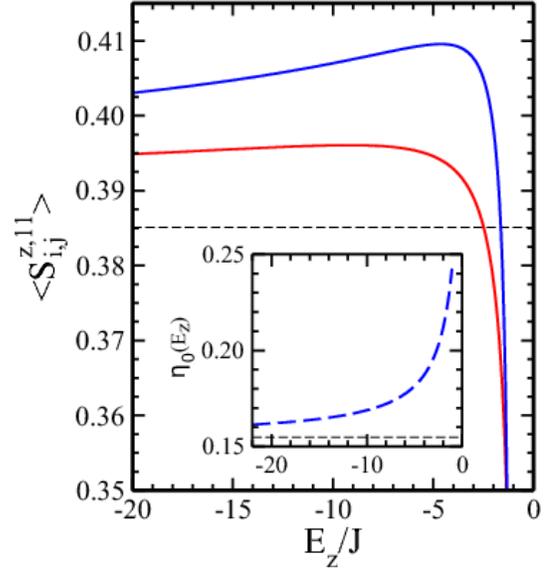}\end{center}
\par\end{centering}
\caption{
The sublattice magnetization $\left\langle S_{i,j}^{z,11}\right\rangle$
as function of $\varepsilon_z$ in the ortho-AF phase for:
$\eta_0$ constant such that $A\left(\eta_0\right)=0$ for 
$\varepsilon_z\to-\infty$ (lower-red curve),
and $\eta_0\left(\varepsilon_z\right)$ such that 
$A\left(\eta_0\right)=0$ for any $\varepsilon_z$ (upper-blue curve). 
The inset shows the function $\eta_0(E_z)$ with its asymptote for 
$\varepsilon_z\to-\infty$ marked with the black dashed line.
}
\label{fig:es_ez} 
\end{figure}

The last step is the Bogoliubov transformation of the block-diagonal
Hamiltonian $H_s$ in the momentum space. The most general form of this 
transformation is: 
\begin{equation}
\left(\begin{array}{c}
\boldsymbol{b}_{\bf k}^{\dagger}\\
\boldsymbol{b}_{-\bf k}
\end{array}\right)=\boldsymbol{{\cal B}}_{\bf k}\!\left(\begin{array}{c}
\boldsymbol{a}_{\bf k}^{\dagger}\\
\boldsymbol{a}_{-\bf k}\end{array}
\right).
\end{equation}
Here $\boldsymbol{b}_{\bf k}^{\dagger}\equiv\{b_{\bf k}^{pq\dagger}\}$ are
the new boson operators and $\boldsymbol{{\cal B}}_{\bf k}$ is an $8\times8$
transformation matrix to be determined from the Bogoliubov-de Gennes 
eigenequation,
\begin{equation}
[H_s,b_{\bf k}^{pq\dagger}]=E_{\bf k}^{pq}\,b_{\bf k}^{pq\dagger},
\end{equation}
equivalent to an $8\times8$ matrix eigenproblem with hyperbolic normalization
conditions typical for bosons. The resulting eigen-vectors form the
rows $\boldsymbol{{\cal B}}_{\bf k}$ and the excitation energies $E_{\bf k}^{pq}$
can be expressed analytically as,
\begin{eqnarray}
\left(E_{\bf k}^{pq}\right)^{2} & = & 2^{4}\left(1-\gamma_{\bf k}^{pq}\right)
\left\{B-6C+\left(B-24C\right)\gamma_{\bf k}^{pq}\right\}\nonumber \\
& \times & \left\{3B+A\gamma_{\bf k}^{pq}-B\left(\gamma_{\bf k}-2
\left(-1\right){}^{p+q}\gamma'_{\bf k}\right)\right\},\nonumber \\
\end{eqnarray}
with
\begin{equation}
\gamma_{\bf k}^{pq}=\frac{1}{2}\left\{\left(-1\right)^{p}\cos\frac{k_{a}}{2}
+\left(-1\right)^{q}\cos\frac{k_{b}}{2}\right\}.
\end{equation}
As might have been expected, in addition to the three gapless Goldstone
modes for 
$\left(p,q\right)=\left(2,2\right),\left(1,2\right),\left(2,1\right)$,
there is one gapped branch $E_{\bf k}^{11}$, with 
$E_{\bf k=0}^{11}=24\sqrt{\left(4B-A\right)C}$,
related to the rigidity of the angle $\varphi$ between the two antiferromagnets.

\begin{figure}[t!]
\begin{centering}
\begin{center}
\includegraphics[width=7cm]{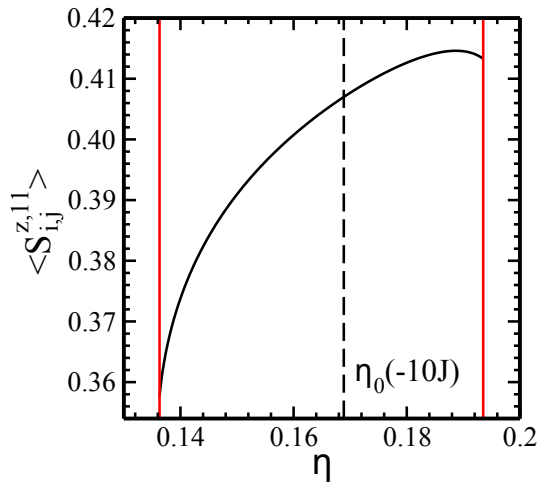}
\end{center}
\par\end{centering}
\caption{
The sublattice magnetization $\left\langle S_{i,j}^{z,11}\right\rangle$ 
as function of $\eta$ in the ortho-AF phase for $\varepsilon_z=-10$ 
(solid line). Dashed line marks the value of $\eta_{0}$ where $A=0$. 
The vertical solid (red) lines are boundaries of the spin-wave 
expansion, where the integral (\ref{eq:Es_int}) becomes divergent.
}
\label{fig:es_et} 
\end{figure}

Another important quantity is the magnetization on a sublattice which
quantifies quantum fluctuations. For instance, the ground state expectation
value of $S_{i,j}^{z,11}$ can be expressed by the elements of 
$\boldsymbol{{\cal B}}_{\bf k}^{-1}$,
\begin{equation}
\left\langle S_{i,j}^{z,11}\right\rangle =\frac{1}{2}-\frac{1}{4\pi^{2}}
\intop_{-\pi}^{\pi}\sum_{p=5}^{8}\left\{
\boldsymbol{{\cal B}}_{\bf k}^{-1}\right\}_{1,p}^{2}d^2k.\label{eq:Es_int}
\end{equation}
The integrand in the above formula can be obtained analytically whereas
the integration is non-algebraic. 
In Figs. \ref{fig:es_ez} and \ref{fig:es_et}
we show the behavior of $\left\langle S_{i,j}^{z,11}\right\rangle$
along different cuts of the ortho-AF phase.

Both cuts in Fig. \ref{fig:es_ez} take the same value of 
$\left\langle S_{i,j}^{z,11} \right\rangle\simeq 0.385$
in the limit of $\varepsilon_z\to-\infty$ when the coupling between 
the two antiferromagnets is negligible. This magnetization is markedly
higher than that for the 2D AF Heisenberg model, where 
$\left\langle S^{z}\right\rangle\simeq 0.303$. This confirms
that the NNN FM interactions in $H_{s}$ make the AF order more 
robust against quantum fluctuations. What is more, along the line 
$\eta_0(\varepsilon_z)$ (blue curve) the third order term $\propto C$ 
enhances the order parameter up to 
$\left\langle S_{i,j}^{z,11}\right\rangle\simeq 0.405$ near 
$\varepsilon_z=-4$. Deviations from the path 
$\eta_0\left(\varepsilon_z\right)$ that introduce nonzero coefficient 
$A$ of the NN AF coupling can either impair or enhance the ortho-AF 
order. For instance, the red curve lies below the blue one in Fig. 
\ref{fig:es_ez}. In contrast, the cut along the line 
$\varepsilon_z=-10$ in Fig. \ref{fig:es_et} shows that a small increase 
of $\eta$ above $\eta_0$ can increase the magnetization (at this value
of $\varepsilon_z$). This reflects the proximity to the ferromagnetic 
phase where the quantum fluctuations reducing the order parameter 
$\left\langle S_{i,j}^{z,11}\right\rangle$ are suppressed.

The sudden collapse of the blue and red curves in Fig. \ref{fig:es_ez} 
terminates the ortho-AF phase in the spin wave approach. 
However, this is not a definitive conclusion because the perturbative 
Hamiltonian $H_s$ underlying the spin wave expansion is 
not self-consistent for $\varepsilon_z>-4$, where the third order term 
$\propto C$ in $H_s$ dominates over the second order term $\propto B$. 
In the non-perturbative plaquette MF the ortho-AF phase extends far 
beyond the perturbative regime, and it is even more extended in the 
more accurate ERA approach.


\begin{thebibliography}{99}
\bibitem{plaq}  This choice is motivated by a particular preference towards
                spin singlets with $|z\rangle$-like orbitals on the bonds.
\bibitem{t_vals} In fact the orbitals are never fully polarized, e.g., considering 
                         finite Hund's exchange $\eta =0.05$ we get
                        $t^c\simeq -0.496$ for  $E_z=-2J$ in the left AF phase,  
                        $t^c\simeq 0.487$ for  $E_z=1J$ in the right AF phase and
                        $t^a\simeq -0.466$ for  $E_z=-0.5J$ in the PVB phase, where
                        the deviations from $t^a=-\frac12$ are typically larger.
\end{thebibliography}
\end{document}